\documentclass[aps,pre,twocolumn]{revtex4}
\usepackage{graphicx}
\usepackage{epsfig}
\usepackage{amsmath}
\usepackage{graphics}
\usepackage{epsfig}
\usepackage{tikz}
\usepackage{epstopdf}
\usepackage{epsfig}
\newcommand{\bra}[1]{\ensuremath{\langle{#1}|\,}}
\newcommand{\ket}[1]{\ensuremath{\,|{#1}\rangle}}

\begin{document}

\title{Distribution of residence times as a marker to distinguish different pathways for quantum transport }
\author{Samuel L. Rudge}
\author{Daniel S. Kosov}
\address{College of Science and Engineering, James Cook University, Townsville, QLD, 4811, Australia }


%%%%%%%%%%%%%%%%%%%%%%%%%%%%%%%%%%%%%%%%%%%%%%%%%%%%%%%%%%%%%%%%%%%%%%
\begin{abstract}
Electron transport through a nanoscale system is an inherently stochastic  quantum mechanical  process. Electric current is   a time series of electron tunnelling events separated by random intervals.   Thermal  and  quantum noise are two sources of this randomness.
In this paper, we used the quantum master equation to consider the following questions:
(i) Given that an electron has tunnelled  into the electronically unoccupied system from the source electrode at some particular time,
how long is it until an electron tunnels out to the drain electrode to leave the system electronically unoccupied, where there were no intermediate tunnelling events (``the" tunnelling path)? 
(ii)  Given that an electron has tunnelled  into the unoccupied system from the source electrode at some particular time, how long is it until an electron tunnels out to the drain electrode to leave the system electronically unoccupied, where there were no intermediate tunnelling events (``an" tunnelling path)? 
(iii)  What are the distributions of these times?  We show that electron correlations  suppress the difference between {\it the}  and  {\it an} electron tunnelling paths.
\end{abstract}

\maketitle

Recently, there have been significant advances towards theoretical and experimental understanding of quantum electron transport through nanoscale systems.
%There are many interesting technological applications for single-molecule electronics but, perhaps, even more important is the indispensability of single-molecular junctions as 
%experimental and theoretical tools to study fundamental problems of open far-from-equilibrium quantum systems. There are huge challenges involved in theoretical modelling of molecular junctions due to all kinds of physical complications - far-from-equilibrium, strong electron-electron and electron-vibration interaction, solvents, complex interface chemistry, importance of different time and length scales, energy and information exchange between quantum systems, and thermal classical environments.
%The relations between molecular structure and conducting properties, crucial role of molecular electrode interfaces\cite{li:035415}, inelastic transport 
%processes \cite{galperin:035301,PhysRevB.80.165305}, role of molecular vibrations\cite{PhysRevB.83.115414,fcblockade05}, coupling of electron transport with  light,  electrochemical gating\cite{solvent12,kuznetsov-ndr},  and  current-induced chemical reactions \cite{PhysRevLett.78.4410,PhysRevLett.84.1527, ho99, Repp26052006, dzhioev11, catalysis12} are among the important problems which are  much better understood now. 
In the past, nanoelectronics research was largely focussed 
on the study of  current-voltage characteristics, the main (macroscopic) observable from which information about microscopic details of electron transport is deduced. However, in recent years, the interest in current fluctuations has grown enormously due to the important   physical information contained within them \cite{nazarov-book}. Additionally, current fluctuations started to play important role in single-molecule electronics determining intricate details of interface chemistry and electron-vibrational coupling \cite{PhysRevB.87.115407,PhysRevB.91.235413,avriller09,thoss14,segal15,PhysRevLett.100.196804,doi:10.1021/nl201327c,doi:10.1021/nl060116e,Tsutsui:2010aa}

The waiting time distribution (WTD) is a natural physical quantity that describes the quantum  transport of single electrons. WTDs for successive physical events have been extensively studied as tools to describe stochastic processes in a diverse range of fields,  from  applied mathematics and  astrophysics to
single-molecule chemistry \cite{Wheatland2002,Baik2006,enzyme-wtd,rna-wtd,cao08}. WTDs were first applied to quantum processes in the 1980s in photon counting quantum optics experiments \cite{Scully1969,Srinivas2010,Vyas1988}.
Recently, WTDs have been used to describe the statistics of single
electron transport in nanoscale quantum systems. 
In 2008 Brandes published his seminal paper on WTDs in quantum transport \cite{brandes08}. His methodology succinctly calculates the distribution
of waiting times for various pairs of electron tunnelling events in
open quantum systems described by general Markovian master equations.
Furthermore, the formalism highlights the connections between WTDs
and other important statistical tools for describing stochastic quantum
processes, such as shot noise, current fluctuations, and full counting
statistics. Recently, Goswami and Harbola used Brandes' approach and the Lindblad master  equation to study stochastic quantum fluctuations  in electron transport through a  single electronic level \cite{harbola15}. 
The theory has
so far mainly focused on distributions of waiting times between events detected by counting
in either the source or collector only \cite{brandes08,buttiker12,flindt13,sothmann14,flindt14,flindt15,wtd-transient}.

In this paper we use Brandes' formalism to calculate the
waiting time distribution between source and drain tunnelling events.
To distinguish this quantity from standard (source-source or drain-drain) WTD, it is called
 the residence
time distribution (RTD) \cite{rudge16a}.
The RTD is defined as the distribution of
times taken for an electron to tunnel across a nanoscale system.
It is characterised by continuous measurements in the source and collector
simultaneously as opposed to counting in just one. 
In order to flow from the source to  the drain  via the nanoscale system, the electron has to navigate through the system Fock space  undergoing (multiple) quantum jumps between state vectors in the process. The wealth of physically interesting information about different tunnelling pathways is revealed by distinct RTDs.
As a model nanoscale system with electronic correlations we consider the Anderson impurity model.
Anderson impurities may be occupied by more than one electron at once, so there are
multiple Fock space paths for tunnelling through the system and hence multiple
RTDs.  
 Inelastic electron scattering processes manifest themselves by suppressing 
some of the tunnelling paths and producing profoundly different RTDs that depend on the strength of electron-electron interactions.
Therefore, the RTD is a useful new tool to study intricate effects of far-from-equilibrium electronic correlations in open quantum systems.
The first moment of the RTD is the average  time delay $\langle\tau\rangle$ between two quantum tunnelling events, which is a quantity important beyond quantum transport \cite{RevModPhys.66.217,RevModPhys.61.917,nitzan00,doi:10.1021/acs.jpcb.5b00862}.

In this paper, we  study the RTD for distinct tunnelling
paths (``the'' and ``an'' electron). {\it The} electron RTD is the conditional probability that an electron is transferred from the nanoscale system to the drain electrode at time $t+\tau$ given that the electron (the same one) was tunnelled to an initially empty system at time $ t$, and that there were no other intermediate tunnelling events. {\it An} electron RTD is the conditional probability that an electron  (any one) is transferred from the system to the drain electrode at time $t+\tau$ given that an electron  was tunnelled to an initially empty system at time $ t$.

% The average tunnelling time is defined
%as the average time between observing an electron entering the 
%from the source electrode and an electron leaving the molecule to the drain electrode.
%Note that this ``average tunnelling time`` definition is one of many, and the topic is highly debated \cite{RevModPhys.66.217,RevModPhys.61.917,nitzan00,doi:10.1021/acs.jpcb.5b00862}.
%The average tunnelling time is
%related to the average current, and to deeper transport statistics
%such as noise and current fluctuations. Therefore, the average
%tunnelling time's dependence on temperature and strength of electronic correlations could provide insight into how these variables affect other transport
%statistics. In this paper we focus on the first moment of the RTD
%and how it is affected by temperature, interaction potential, and
%tunnelling paths.

%The paper is organised as follows. Section II describes the Markovian
%master equation for an Anderson impurity coupled to two electrodes, as well as identifies the single electron  jump superoperators
%in the system Liouvillian. In Section III, we apply the theory to
%an Anderson impurity and calculate the RTD for various tunnelling
%paths (``the'' and ``an'' electron). Additionally, we calculate
%the temperature and interaction potential dependence of the average
%waiting time $\langle\tau\rangle$. Section IV summarises the main
%results of the paper. 

We use natural units in equations throughout the paper: $\hbar=k_{B}=e=1$.

Let us consider a nanoscale quantum system  linked to two macroscopic metal electrodes, the source and drain, which are held at different chemical potentials.
The total Hamiltonian is
\begin{equation}
 {H}= {H}_{S}+ {H}_{D}+ {H}_{Q}+ V.
\end{equation}
The source and drain electrodes contain  non-interacting electrons and are described by the following Hamiltonians:
\begin{equation}
  {H}_{S}=\sum_{s \sigma} \varepsilon_{s}a_{s\sigma }^{\dagger}a_{s\sigma} , \;\;\;\;\  {H}_{D}=\sum_{d\sigma }\varepsilon_{d}a_{d\sigma }^{\dagger}a_{d\sigma}.
 \end{equation}
 Here, $a^\dagger_{s\sigma/d \sigma}$   creates  an electron with spin $\sigma=\uparrow, \downarrow$ in the single-particle state $s/d$ of the source/drain electrode and $a_{s\sigma/d\sigma}$ is the corresponding electron  annihilation operator.
The quantum systems is described  by the  Anderson model Hamiltonian:
\begin{equation}
H_{Q}=\epsilon\sum_{\sigma}a_{\sigma}^{\dagger}a_{\sigma}+Ua^\dag_{\uparrow} a_{\uparrow} a^\dag_{\downarrow} a_{\downarrow},
\label{hm}
\end{equation}
where the
operator $a^\dag_\sigma (a_\sigma)$ creates (destroys) an electron with spin $\sigma$ on the single-particle level with energy $\epsilon$  and $U$ is the electron-electron repulsion.
The tunnelling coupling between the system and electrodes is
 \begin{align}
V= t_S \sum_{s\sigma} (a_{s \sigma}^{\dagger}a_{\sigma }+  a^\dag_{\sigma }a_{s \sigma} ) + t_D \sum_{d\sigma}  (a_{d \sigma}^{\dagger}a_{\sigma }+ a^\dag_{\sigma }a_{d \sigma} ).
\label{v}
\end{align}

The Fock space of the quantum system consists of 4 states. The system can either be empty, occupied by a spin up or a spin down electron,
or occupied by both a spin up and a spin down electron:
$ |0\rangle$,  $\ket{\uparrow}=a_{\uparrow}^{\dagger}|0\rangle$,   
$\bra{\downarrow}=a_{\downarrow}^{\dagger}|0\rangle $, and  
$|2 \rangle=a_{\uparrow}^{\dagger}a_{\downarrow}^{\dagger}|0\rangle $. 
\begin{widetext}
Using these states as a  basis, we write the master equation for the  system density matrix:
\begin{eqnarray}
\frac{d}{dt}
\begin{bmatrix}
{\bra{0} \rho \ket{0}}\\
{\bra{\uparrow} \rho \ket{\uparrow}}\\
{\bra{\downarrow} \rho \ket{\downarrow}}\\
{{\bra{2} \rho \ket{2}}}
\end{bmatrix} & = & \left[\begin{array}{cccc}
-2T_{01} & T_{10} & T_{10} & 0\\
T_{01} & -\Big(T_{10}+T_{12}\Big) & 0 & T_{21}\\
T_{01} & 0 & -\Big(T_{10}+T_{12}\Big) & T_{21}\\
0 & T_{12} & T_{12} & -2T_{21}
\end{array}\right]
\begin{bmatrix}
{\bra{0} \rho \ket{0}}\\
{\bra{\uparrow} \rho \ket{\uparrow}}\\
{\bra{\downarrow} \rho \ket{\downarrow}}\\
{{\bra{2} \rho \ket{2}}}
\end{bmatrix}.
\end{eqnarray}
\end{widetext}
Here $\rho$ is the system density matrix and $T_{nm}$ is the rate of transforming from a  state occupied by $n$ electrons to state populated by $m$ electrons.

 The rates for electron  transfer are computed using the Fermi golden rule with the tunnelling interaction (\ref{v})  treated as a perturbation.
The tunnelling rates for electron transfer between the source electrode and the system are
\begin{eqnarray}
T^{S}_{01}=\Gamma_{S}f_{S}({\epsilon}), \;\;
T^{S}_{10}=\Gamma_{S}(1-f_{S}({\epsilon})),\\
T^{S}_{12}=\Gamma_{S}f_{S}({\epsilon+ U}), \;\;
T^{S}_{21}=\Gamma_{S}(1-f_{S}({\epsilon+ U})).
\end{eqnarray}
Likewise, for tunnelling between the drain electrode and the system the tunnelling rates are\
\begin{eqnarray}
T^D_{01}=\Gamma_{D}f_D({\epsilon}), \;\;
T^D_{10}=\Gamma_{D}(1-f_D({\epsilon})),\\
T^D_{12}=\Gamma_{D}f_D({\epsilon+ U}), \;\;
T^D_{21}=\Gamma_{D}(1-f_D({\epsilon+ U})).
\end{eqnarray}
where $ \Gamma_{S/D} =2 \pi   |t_{S/D}|^2  \rho_{S/D}$ and $ \rho_{S/D}$ is the density of states for the source/drain electrodes. 
The functions $f_S$ and $f_D$ are Fermi occupation numbers for the source and drain electrodes:
\begin{equation}
f_S(\epsilon)=[1+e^{(\epsilon-\mu_S)/T_S}]^{-1}, \;\;\;\; f_D(\epsilon)= [1+e^{(\epsilon-\mu_D)/T_D}]^{-1}.
\end{equation}
The voltage bias is defined as the difference between the source and drain chemical potentials
$\mu_S -\mu_D$.
The total rates combine contributions from the source and drain electrodes:
\begin{eqnarray}
T_{10}=T_{10}^{S}+T_{10}^{D}, \;\;\;\;T_{01}=T_{01}^{S}+T_{01}^{D},
\\
T_{21}=T_{21}^{S}+T_{21}^{D},
 \;\;\;\;
T_{12}=T_{12}^{S}+T_{12}^{D}.
\end{eqnarray}

Introducing probabilities that the system is empty $P_0= {\bra{0} \rho \ket{0}} $, occupied by one electron $P_1={\bra{\uparrow} \rho \ket{\uparrow}} + {\bra{\downarrow} \rho \ket{\downarrow}}$, and occupied by two electrons 
$P_2={\bra{2} \rho \ket{2}}$, the master equation becomes
\begin{eqnarray}
\frac{d}{dt}
\begin{bmatrix}
{P_0}\\
{P_1}\\
{P_{2}}
\end{bmatrix} 
& = & L
\begin{bmatrix}
{P_0}\\
{P_1}\\
{P_{2}}
\end{bmatrix},
\label{master}
\end{eqnarray}
where the Liouvillian is
\begin{eqnarray}
L
& = & \left[\begin{array}{cccc}
-2T_{01} & T_{10} &  0\\
2T_{01} & -\Big(T_{10}+T_{12}\Big)  & 2 T_{21}\\
0 & T_{12} &  -2T_{21}
\end{array}\right].
\label{L}
\end{eqnarray}

%We define a quantum jump operator as a super-operator that, upon acting on a pure state, instantly maps the density 
%matrix to another pure state. These operators model single electrons `jumping' into and out of the molecule via quantum tunnelling, hence the name. 
%\begin{eqnarray}
%	J \rho&=&\rho'
%\end{eqnarray}
%Here, $J$ is a quantum jump operator and $\rho,\rho'$ are density matrices describing different molecular states.
%Consequently, quantum jumps will always occur on the off-diagonal
%elements of the Liouvillian $L$, as they are the only elements positively contributing
%the rate of change of the pure state under consideration.

We formally split the  Liouvillian $L$ into a diagonal part $L_{0}$  with the remainder described by a sum of quantum jump operators
$J_{nm}^{S}$ and $J_{nm}^{D}$,  each of which transforms a  state  occupied by $n$ electrons to a state with $m$ electrons, through an interaction with the source and drain electrodes respectively. The Liouvillian is
\begin{eqnarray}
L  =  L_{0} \\
 + J_{01}^S+J_{10}^S + J_{21}^S+J_{12}^S +J_{01}^D+J_{10}^D + J_{21}^D+J_{12}^D,
\nonumber
\end{eqnarray}
where
\begin{eqnarray}
L_0
& = & \left[\begin{array}{cccc}
-2T_{01} & 0 & 0\\
0 & -\Big(T_{10}+T_{12}\Big) &0 \\
0 & 0 &  -2T_{21}
\end{array}\right],
\end{eqnarray}
and each jump operator is a 3x3 matrix with a single non-zero element.

The jump operator $J_{01}^{S/D}$  transforms the system from the empty  state to 
the singly occupied state by tunnelling of a spin up or down electron from the source/drain electrode into the system:
\begin{eqnarray}
\label{j01}
J_{01}^{S/D}
 = |J_{01}^{S/D})(\widetilde J_{01}^{S/D}|
 = \left[\begin{array}{c}
0 \\
1 \\
0
\end{array}\right] \left[ 2T_{01}^{S/D} \;\;   0  \;\; 0 \right].
\end{eqnarray}
The jump operator $J_{10}^{S/D}$ describes the reverse process: it  transforms the system from being occupied by one electron to being 
empty by transferring one electron from the system to the source/drain electrode:
\begin{eqnarray}
\label{j10}
J_{10}^{S/D}
 = |J_{10}^{S/D})(\widetilde J_{10}^{S/D}|
= \left[\begin{array}{c}
1\\
0 \\
0
\end{array}\right]
\left[ 0 \;\;   T_{10}^{S/D} \;\; 0 \right].
\end{eqnarray}
The jump operator $J_{12}^{S/D}$  transforms the system from being occupied by one electron to two
by transferring one electron of opposite spin from the source/drain electrode to the system:
\begin{eqnarray}
\label{j12}
J_{12}^{S/D}
 = |J_{12}^{S/D})(\widetilde J_{12}^{S/D}|
 =  \left[\begin{array}{c}
0 \\
0 \\
1
\end{array}\right] \left[  0\;\;   T_{12}^{S/D}  \;\; 0 \right] .
\end{eqnarray}
Conversely, the jump operator $J_{21}^{S/D}$  transforms the fully occupied system to being occupied by a single, either spin up or down, electron
by transferring one electron from the system to the source/drain electrode
\begin{eqnarray}
\label{j21}
J_{21}^{S/D}
 = |J_{21}^{S/D})(\widetilde J_{21}^{S/D}|
= \left[\begin{array}{c}
0\\
1 \\
0
\end{array}\right]
\left[ 0 \;\;  0 \;\;  2 T_{21}^{S/D} \right].
\end{eqnarray}

Now,
we are ready to define  the  RTD.
First,  we need to decide which quantum jumps we would like to monitor directly and which  jumps will be run as ``background processes" during the time evolution of the density matrix:
\begin{equation}
\label{L1}
 {L}  =   {\mathcal L}_{0}+ \sum_{\text{monitored}} J_i.
\end{equation}

Consider  the system in the  nonequilibrium steady state $\overline \rho$:
\begin{equation}
L \overline \rho =0.
\end{equation}
 Suppose that we observe an electron tunnelling from the source to the nanoscale system, $J_\alpha^{S}$ at $t_1$ ($\alpha=\{01\}$ if the system was unoccupied before the electron transfer, or $\alpha=\{12\}$ if the electronic occupation is increased from 1 to 2 by electron tunnelling), and then
 we observe another quantum tunnelling from the system to the drain electrode  $J_\beta^{D}$  ($\beta=\{10\}$ or $\beta=\{21\}$) at some later  time  $t_2$.
 The joint probability distribution that the system undergoes quantum jump  $J_{\alpha}$ at  $t$
 and  then another jump  $J_{\beta}$  in time   $t +\tau$:
\[
P(\tau) = \text{Tr} [ J^D_{\beta}e^{ {L}_{0}\tau} J^S_{\alpha} \overline\rho].
\]
Since the probability  to have quantum jump $J^S_{\alpha}$ in arbitrary time after the establishment of the steady state is
\[
p = \text{Tr} [ J^S_{\alpha} \overline\rho],
\]
we rewrite $P(\tau)$ as
\[
P(\tau) = \frac{\text{Tr} [  J^D_{\beta}e^{ {L}_{0}\tau} {J}^S_{\alpha} \overline\rho]}{\text{Tr} [ J^S_{\alpha} \overline\rho]} \;  \text{Tr} [ J^S_{\alpha} \overline\rho].
\]
Comparing the last two equations with  the standard Kolmogorov relation between joint and conditional probabilities, we identify the expression for  the RTD,  $w_{\alpha \beta} (\tau)$, as the conditional probability that the system undergoes $J^D_\beta$ at time $t_1 + \tau$ given 
that it underwent $J^S_\alpha$ at time $t_1$:
\begin{equation}
w_{\alpha \beta}(\tau) = \frac{\text{Tr} [  J^D_{\beta}e^{ {\mathcal L}_{0}\tau} J^S_{\alpha} \overline\rho]}{\text{Tr} [ J^S_{\alpha} \overline\rho]}. 
\label{w}
\end{equation}
Using superoperator techniques \cite{brandes08,dzhioev11a, dzhioev11b,dzhioev12,dzhioev14,dzhioev15,harbola08},  we get
\begin{equation}
w_{\alpha \beta} (\tau) = (  {\tilde J}^D_{\beta}
|e^{ {\mathcal L}_{0}\tau} |  {J}^S_{\alpha}),
\label{w-final}
\end{equation}
where $ (  {\tilde J}^D_{\beta}$ and $ |  {J}^S_{\alpha})$ are supervectors defined via dyadic representation of quantum jump operators (\ref{j01}-\ref{j21}). For a detailed derivation of  the expression for the RTD (\ref{w-final}) we refer to  \cite{brandes08,rudge16a}.
We note that (\ref{w-final}) is less general than the starting general expression for the RDT (\ref{w}) since its derivation requires the representation of jump operator (which is a second order tensor) as a  single dyadic product \cite{rudge16a}.

We focus on the RTD
which is the conditional probability that an electron is transferred from the system to the drain electrode at time $t + \tau$ given 
that an electron tunnelled into an initially empty system at time $t$. It is defined as
\begin{eqnarray}
 w(\tau)  && =  ( {\widetilde J}^D_{10}|e^{ {\mathcal L}_{0}\tau} |  J^S_{01}) 
 =  
\left[ 0 \;\;   T_{10}^{D} \;\; 0 \right]  e^{ {\mathcal L}_{0}\tau} 
\left[\begin{array}{c}
0 \\
1 \\
0
\end{array}\right] 
\nonumber
\\
&& = T_{10}^{D} [e^{ {\mathcal L}_{0}\tau}]_{22},
 \label{ttd}
\end{eqnarray}
where $[e^{ {\mathcal L}_{0}\tau}]_{22}$ is the element in row $2$ and column $2$ of  matrix $e^{ {\mathcal L}_{0}\tau}$.

The physical interpretation of this RTD  (\ref{ttd}) depends on the definition of $ {\mathcal L}_{0}$, which is determined by the splitting of the total Liouvillian into monitored quantum jumps and the generator for background quantum dynamics. We will consider two cases - {\it the} electron and {\it an} electron tunnelling. 

Suppose that we monitor all possible quantum jumps in the system
\begin{equation}
\label{full}
L  =  \underbrace{L_{0}}_{\mathcal L_0} + \underbrace{J_{01}^S+J_{10}^S + J_{21}^S+J_{12}^S +J_{01}^D+J_{10}^D + J_{21}^D+J_{12}^D}_{\text{monitored}} 
\end{equation}
In the intermediate time $\tau$, the system is forced to evolve without the removed quantum jumps due to $e^{\mathcal{L}_0\tau}$. In this case every quantum jump has been removed from $\mathcal{L}_0$, so that after ${J}^S_{\alpha}$ occurs the RTD examines only the probability distribution for when there are no intermediate quantum jumps until ${J}^D_{\beta}$ occurs sometime later at $t+\tau$. We label this tunnelling path as {\it the} electron, because it describes the distribution of waiting times between consecutive tunnelling events with {\it the} same electron. Note that we could not draw the same conclusion for quantum jumps involving multiply occupied systems (e.g. $J^S_{12}$ and $J^D_{21}$) due to electron indistinguishability. The consecutive pathway is:

\[
|0\rangle\langle0|\overset{J_{01}^{S}}{\rightarrow}|1\rangle\langle1|\overset{J_{10}^{D}}{\rightarrow}|0\rangle\langle0|
\]

In contrast, suppose that we monitor only two tunnelling events: the transfer of an electron to an initially empty system from the source electrode, $J_{01}^S$, and tunnelling of {\it an} electron (any one) from a singly occupied system to the drain electrode, $J_{10}^D$. We use the same distribution as (\ref{ttd}). 
 The rest of the quantum jumps are included into $\mathcal L_0$:
\begin{equation}
L  =  \underbrace{L_{0}+J_{10}^S + J_{21}^S+J_{12}^S +J_{01}^D + J_{21}^D+J_{12}^D}_{ \mathcal L_0} + \underbrace{ J_{01}^S + J_{10}^D }_{\text{monitored}}
\end{equation}
Here, the time-evolution $\mathcal{L}_0$ between the two monitored tunnelling events contains the other quantum jump operators. After the initial quantum jump $J_{01}^S$ the system may undergo intermediate quantum jumps in $\tau$ before the quantum jump $J_{10}^D$. We label this tunnelling pathway as {\it an} electron, because it describes the distribution of waiting times between any two electrons. An example tunnelling pathway is:

\[
|0\rangle\langle0|\overset{J_{01}^{S}}{\rightarrow}|1\rangle\langle1|\overset{J_{12}^{S}}{\rightarrow}|2\rangle\langle2|\overset{J_{21}^{D}}{\rightarrow}|1\rangle\langle1|\overset{J_{10}^{D}}{\rightarrow}|0\rangle\langle0|
\]

Note that this is only an example tunnelling pathway for {\it an} electron. The number of {\it an} electron tunnelling pathways is technically infinite, and they include {\it the} electron tunnelling pathway.

\begin{figure}[t!]
\begin{center}
\includegraphics[width=\columnwidth]{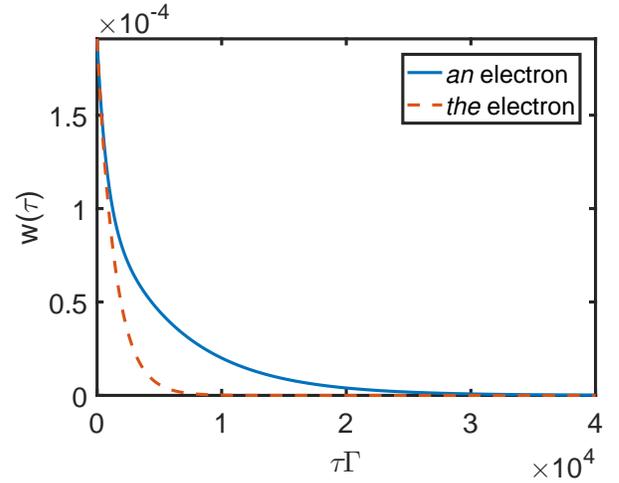}
\end{center}
	\caption{RTDs $w(\tau)=(\tilde{J}_{10}^{D}|e^{\mathcal{L}_{0}\tau}|J_{01}^{S})$ for `an' and `the' electron. Parameters  used in calculations are: $\epsilon=-0.3\Gamma$, $U=0.5\Gamma$, $\mu_{S/D}=\pm0.5\Gamma$,  and $T=2.585\Gamma$.}	
\label{wtd}
\end{figure}

The RTDs for \textit{the} electron and \textit{an} electron are shown in Fig.\ref{wtd}. as a function of residence time  $\tau$.  The distributions are exponentially decaying, and \textit{an} electron's RTD is characteristically long-tailed compared to \textit{the} electron. This corresponds to a longer average residence time $\langle\tau\rangle$, which is intuitively expected due to the presence of  intermediate quantum jumps. 

\begin{figure}[t!]
\begin{center}
\includegraphics[width=\columnwidth]{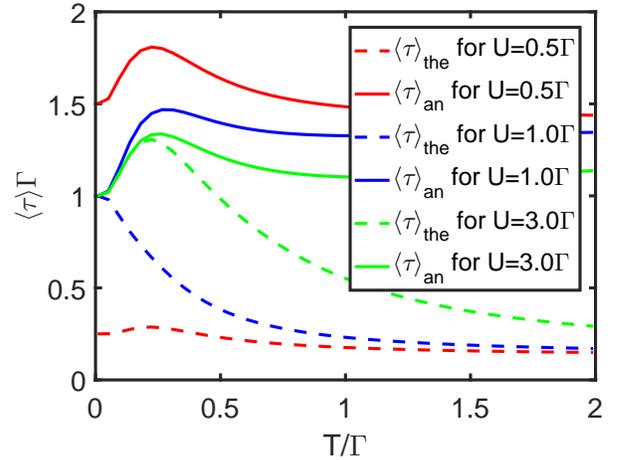}
\end{center}
	\caption{Temperature dependence of the average residence time computed for various values of electron-electron repulsion $U$. Parameters used in the calculation are: $\epsilon=-0.3\Gamma$, and $\mu_{S/D}=\pm0.5\Gamma$.}	
\label{homo}
\end{figure}

Let us now discuss how temperature influences the residence times.
The average residence times can be computed analytically. The average residence time for  \textit{the}  electron transport is
\begin{eqnarray}
\langle\tau\rangle_{the}  =  \frac{T_{10}^{D}}{[T_{10}+T_{12}]^{2}},\label{eq:average_tau}
\end{eqnarray}
 whereas the average residence time for {\it an} electron transport is $\langle\tau\rangle_{the}$ multiplied by a pre-factor:
\begin{equation}
\langle\tau\rangle_{an}  = \frac{[T_{10} + T_{12}]^2 [T_{01}^D T_{10}^S T_{21} + T_{01}^2 (T_{12} + 2 T_{21})]}{2 [T_{01} T_{10} - 
   T_{01}^D T_{10}^S]^2 T_{21}}\;  \langle\tau\rangle_{the}.
   \label{tau_an}
\end{equation}
The pre-factor in (\ref{tau_an}) is always greater than or equal to one.
Intuitively, $\langle\tau\rangle_{an}$ is expected to
be always larger  or equal than $\langle\tau\rangle_{the}$ due to quantum jumps $J_{10}^S$, $J_{01}^D$,   $J_{12}^{S/D}$, and $J_{21}^{S/D}$ occurring  as intermediate
tunnelling events.  If we let the temperature go to zero and assume that the electron-electron repulsion is very large ($U \rightarrow \infty$ limit), then $ \langle\tau\rangle_{an} = \langle\tau\rangle_{the}$. This is again intuitively expected, since the backscattering events  $J_{10}^S$, $J_{01}^D$ are subdued at  low temperature and the jumps involving a doubly populated system $J_{12}^{S/D}$, $J_{21}^{S/D}$  become negligible at large $U$.

This behaviour is shown in Fig.\ref{homo}. 
The initial increase of the residence time at small temperature, so that electrons spend longer time in the nanoscale system before jumping out to the drain, is due to blocking of the available states in the drain electrode by spreading the thermal Fermi-Dirac distribution.
As the temperature increases further, the electronic population of the nanoscale system starts to grow  and the  intermediate quantum jumps  $J_{12}^{S},J_{21}^{S},$ and $J_{21}^{D}$ start to play larger roles.  Quantum jumps involving a doubly populated system   $J_{12}^{S},J_{21}^{S},$ and $J_{21}^{D}$ influence the $\langle\tau\rangle_{the}$  by scaling the RTD up or down via the normalisation. Although these quantum jumps 
are not present directly in (\ref{ttd}), they are involved in the calculations of the normalisation: the RTD is normalised to unity if it is integrated over time from 0 to $\infty$ and is summed over all possible secondary quantum jumps.  This is reflected in the monotonic decrease of average residence times at larger temperature. When $U=0.5\Gamma$ $\langle\tau\rangle_{the}$ and $\langle\tau\rangle_{an}$ are different at all temperatures, as all $\epsilon+U<\mu_S$ and quantum jumps involving a doubly populated system are energetically allowed at all temperatures. At $U=1\Gamma$, both average residence times are the same at small temperature, as $\epsilon+U>\mu_S$ and the quantum jumps involving a doubly populated system are energetically suppressed. As temperature increases they diverge from each other, due to the spreading thermal Fermi-Dirac distribution and the width of the energy level $\Gamma$. A similar behaviour is shown for the larger electron-electron interaction $U=3\Gamma$, except the average residence times start to deviate  at higher temperatures.

Let us now discuss the possible experimental measurements of the RTD in electron transport through nanoscale quantum systems. Suppose that we have an auxiliary detection system that consists of three quantum dots  (QD1, QD2, and QD3), where each is attached to their own source and drain electrodes. The electric current in these quantum dots can be monitored experimentally. QD1 and QD2  are capacitively coupled to the source-system and system-drain barriers, respectively,  to monitor tunnelling events and the QD3 capacitively measure the charging state of the system. Capacitive coupling enables minimally-invasive extraction of information about system charging states and tunnelling events. The QD3 acts as an electrometer by detecting the discrete charge on the SET, thus informing on which of the three charging states the system is in: $0,1$, or $2$. Without QD1 and QD2 this would exactly corresponds to the experimental setup used in \cite{Lu2003} . However, if two more quantum dots (QD1 and QD2) are capacitively coupled to the source-system and system-drain barriers, then single-electron tunnelling events can be detected. We propose the following measuring protocol. The QD3 measures the system to be in the unoccupied state state, then QD1 measures a jump from the source to the system, changing it to state with one electron. Suppose that after some time $\tau$  QD2 detects a tunnelling event from the system to drain, leaving the system in the unoccupied state. If this secondary tunnelling event was detected before another tunnelling detection from QD1, then the time $\tau$ is used in ``the" electron RTDs. If intermediate tunnelling events are detected from QD1, then the time $\tau$ is used in ``an" electron RTD only.

In conclusion, we have developed a  Markovian master equation based theory for RTDs through Anderson impurity model. The theory provides the opportunity to break down electron transport into several paths in the system Fock space. This partitioning  of macroscopic electric current is a practical tool for physically interpretating microscopic mechanisms of electron transport.
To illustrate the general theory we  studied the RTD for two tunnelling
paths (``the'' and ``an'' electron). {\it The} electron RTD is the conditional probability that an electron is transferred from the system to the drain electrode given that an electron (the same one) was tunnelled to an initially empty system at an earlier time. {\it An} electron RTD is the conditional probability that any electron is transferred from the system to the drain electrode given that an electron initially tunnelled into the system at time $t$. We demonstrated  that {\it an} electron has a characteristically longer distribution than {\it the} electron. Furthermore, the two transport pathways have significantly different temperature dependences, which is sensitive to the strength of electron-electron interaction. Inelastic electron-electron  scattering processes manifest themselves by suppressing the difference between {\it the}  and  {\it an} electron tunnelling paths.  Therefore, the RTD is a useful  new theoretical method to unravel intricate details of  inelastic electron transport processes.

The authors thank D. Cocks and M. Gelin for many valuable discussions.

\nocite{}

\end{document}